\documentstyle[12pt]{article}
\begin{titlepage}   
\title{ 
{
\vspace{-8.5ex}
\normalsize
\begin{flushright} 
IHEP 00--02\\
\end{flushright}
\vspace{10ex}
}
{\bf Parameter counting}\\
{\bf for neutrino mixing}\\[3ex]
}
                                                                     
\author{ Yu.~F.~Pirogov{}\thanks{E-mail: pirogov@mx.ihep.su}
\\[1ex]
{\it Institute for High Energy Physics,}\\
{\it Protvino, RU-142284 Moscow Region, Russia}\\[0.5ex]
{\it  Moscow  Institute  of  Physics  and  Technology,}\\
{\it  Dolgoprudny,  Moscow  Region, Russia }
}
\date{}

\begin{document}
\maketitle
\abstract{
\noindent
The content of  physical masses, mixing angles
and $CP$ violating phases in the lepton sector of   extended
standard model, both renormalizable and non-renormalizable,  with
arbitrary numbers of the singlet and left-handed
doublet neutrinos is systematically analysed in the weak basis.
}
\vspace{8ex}
\thispagestyle{empty}
\end{titlepage}

\setcounter{page}{2}
\section{Introduction}

The mixing of  quarks in the minimal Standard Model (SM)
of  strong  and  electroweak interactions 
is now well understood. It is described by the
Cabibbo-Kobayashi-Maskawa (CKM) unitary mixing
matrix for the charged currents~\cite{koba}, the neutral ones and
Yukawa interactions being flavour conserving. 
As for the lepton sector, the SM exhibits here an extremely
simple and economic structure. It encounters
just three physical parameters, the  charged lepton masses, and
predicts no flavour and $CP$ violation. 
But  it has been widely
recognized that the inclusion of  (iso)singlet
neutrinos and/or neutrino masses in the SM  would result in 
the  lepton mixing and flavour violation with all the related
phenomena such as  $CP$ violation, neutrino
oscillations~\cite{ponte}, etc.\ (as a recent review see, e.g.,
ref.~\cite{now}).

There are two principal differences between the lepton and quark
mixings. First, the number of singlet neutrinos relative to that
of (iso)doublet ones is not restricted by the chiral anomalies and
can be arbitrary.
Second, the Majorana masses for neutrinos are possible in addition to
the
Dirac ones.  
This complicates inevitably  the proper 
SM extensions and proliferates  free parameters. 
Hence, an immediate  problem arises  how to extract  the physical
parameters, to 
separate masses,  mixing  angles and $CP$ violating
phases among them, as well as to conveniently 
parametrize the mixing matrices. 
There have been many studies of the related topics. The case with an
arbitrary number of the left-handed doublet neutrinos but without
singlet ones was considered in~ref.~\cite{n0}, the one  with
equal arbitrary numbers of the singlet and doublet neutrinos
in ref.~\cite{nn} and  a  general
case with arbitrary numbers of  both types of neutrinos in
ref.~\cite{nm}. In particular,  the last case with
only  Dirac masses was studied in ref.~\cite{dir}. Traditionally,
all the previous investigations were carried out by an explicit
construction in the mass basis.

In  the present paper an  alternative approach to the lepton 
parameter counting  is adopted. It is formulated in the weak basis
entirely through symmetry properties  of the model
before the spontaneous symmetry breaking~\cite{santa}. In these
terms, all the possible parameter space configurations
of the SM extended with arbitrary numbers of
the singlet and left-handed doublet neutrinos  are 
systematically
analysed. Both renormalizable and non-renormalizable
extensions of the SM, among them the pure Dirac and pure 
Majorana cases, are considered. In a consistent fashion, the known
results on the lepton parameter counting for  the SM
general extensions are recovered, and the new ones for the
renormalizable extensions are obtained. The relation between the two
countings is clarified. The results  on  parameter counting for
neutrino mixing are summarized in tables in what
follows.\footnote{In fact, what we are talking about is
lepton mixing which is described
by a counterpart of the CKM matrix. But one can
always choose a weak basis where the mixing matrix of charged leptons
is unity. In this sense, lepton mixing is synonimous with the neutrino
one.} 

\section{Renormalizable  extensions}
\paragraph{(i) Arbitrary case} 
The most general renormalizable
$SU(2)_{\mbox{\scriptsize W}}\times
U(1)_{\mbox{\scriptsize Y}}$ invariant  lepton
Lagrangian of the SM extended by  the right-handed neutrinos  reads
\def\d{\partial\hspace{-1.2ex}/\hspace{0.3ex}}
\def\D{D\hspace{-1.5ex}/\hspace{0.5ex}}
\begin{eqnarray}\label{eq:lagrangian}
{\cal L}&=&
~~~ \overline{ l_L}i\D  l_L +  \overline{
e_R}i\D
e_R 
  +  \overline{ \nu_R}i\d  \nu_R\nonumber\\
&& - \Big{(} \overline{ l_L} Y^e  e_R\phi 
           + \overline{l_L} Y^\nu  \nu_R \phi^C
           + \frac{1}{2} \overline{ \nu_L^{C}} {M}
           \nu_R
           + \mbox{h.c.}
     \Big{)}\,. 
\end{eqnarray}
In eq.~(\ref{eq:lagrangian}) the   
lepton doublet $l_L$  and singlet $e_R$, $\nu_R$
fields mean those in a weak basis
where, by definition, the symmetry properties are well stated.
It is supposed that  the ordinary chiral families of the SM
with the doublet left-handed   Weyl neutrinos in
number $d\ge 3$  are added by the singlet  Weyl
neutrinos in number $s\geq 0$.  Let us designate the SM extended in
such a renormalizable manner as $(d,s)_{\mbox{\scriptsize r}}$ 
extension. A priori, one should  retain $s$ and
$d$ as arbitrary integers, both  $s\leq d$ and  $s>d$ being
allowed.\footnote{We omit the possible
vector-like lepton doublets in the present analysis. Hence, with
account for the most
probable exclusion
of the fourth heavy chiral family~\cite{4f}, one should put in
reality $d=3$. Nevertheless, we
retain $d$ as a free parameter to better elucidate the parameter
space structure of the extended SM.}   Further, 
$\D\equiv\gamma^\alpha D_\alpha$ is the generic covariant derivative
which reduces to the ordinary one, $\d =
\gamma^\alpha\partial_\alpha$, 
for the hypercharge zero singlet neutrinos.
Here and in what follows the notations  ${ \nu}^{C}_L
\equiv
({\nu}_R)^C  = C\overline{{\nu}_R}\,{}^T$, etc,
are used for the $C$ conjugates of chiral fermions. $Y^e$ and
$Y^\nu$ are the arbitrary complex $d\times d$ and
$d\times s$ Yukawa  matrices, respectively, and 
$M$ is a complex symmetric $s\times s$ matrix of the Majorana masses
for the singlet neutrinos.
Finally,  $\phi$ is the Higgs isodoublet and $\phi^C\equiv i\tau_2
\phi^*$ is its charge conjugate.

\begin{table}[htbp]
\paragraph{Table 1} Parameter counting 
for the SM renormalizable extension
$(d,s)_{\mbox{\scriptsize r}}$  with
$d$ doublet neutrinos and $s$ singlet ones.  In this table  and  the
ones
which follow, the first and the second groups of moduli for the
physical mass matrix ${\cal M}_{ph}$ correspond to the independent
mixing angles and masses, respectively (see text).
\vspace{1ex}
\begin{center}
\begin{tabular}{|c|c|c|}
\hline 
Couplings&Moduli&Phases\\
and symmetries&&\\
\hline
$Y^e$,  $Y^\nu$, $M$ &$d^2+ds +s(s+1)/2$&
$d^2+ds +s(s+1)/2$\\ 
\hline
$G = U(d)^2\times U(s)$&$- d(d-1) -  s(s-1)/2 $& 
$- d(d+1)-  s(s+1)/2 $\\
\hline
$H = I$&$0$&$0$\\
\hline
${\cal M}_{ph}(d,s)_{\mbox{\scriptsize r}}$ &
$sd+(d+s)$& $d(s-1)$\\
\hline
${\cal M}_{ph}(n,n)_{\mbox{\scriptsize r}}$&
$n^2+2n $&$n(n-1)$\\
\hline
\end{tabular}
\end{center}
\end{table}

The parameter counting in the weak basis  for the
lepton sector of the extended SM  proceeds as
is shown in  Table~1. Here $G$ is the global symmetry of the kinetic
part of the Lagrangian of eq.~(\ref{eq:lagrangian}). 
Due to the Dirac and Majorana mass terms the symmetry $G$
is  explicitly violated  so that the  residual symmetry is trivial, 
$H=I$.\footnote{In what folows, we generally assume that there are no
mass textures or accidental mass degeneracy. Otherwise, the residual
symmetry would increase, and  special consideration of each
particular case would be mandatory.}  The transformations of the
broken part $G/H$ (here $G/H=G$) can be used to
absorb the spurious  parameters in eq.~(\ref{eq:lagrangian}) leaving
only the independent physical set, ${\cal M}_{ph}$, of
them.\footnote{For this
reason, parameters corresponding to symmetry $G$ are represented in
tables with a minus sign, whereas those of $H$
with a plus sign.} As a result, ${\cal M}_{ph}$ contains $sd+d+s$
independent moduli and $d(s-1)$ phases. In this, real ${\cal M}_{ph}$
corresponds to $CP$ conservation.  Stress that only the total number
of independent physical moduli is fixed by the weak basis counting. 
Due to absence of the left-handed Majorana
masses in the weak basis, there are  relations in the
$(d,s)_{\mbox{\tiny r}}$  extensions between the actual mixing
angles and masses. Considering all the masses as independent ones,
while a part of mixing angles as a function of the masses, would
result in the formal number of mixing angles less than
their actual number.  This may cause some confusion in explicit
parametrization.
So, it is more instructive to choose all the  mixing
angels as independent ones, considering part of the masses as a
function of the angles and the rest of the masses. To decide  what
is the minimal number of independent masses, consider the limit
$M\to\infty$ corresponding to decoupling of $s$ heavy
Majorana neutrinos. In this limit, the
rest of $d$ Majorana neutrinos should become massless with necessity.
Thus $d$ Majorana masses depend on $s$ ones. Clearly, it is
impossible to further reduce the number of independent masses.
Finally, of the physical
moduli,  $sd$ ones are  mixing angles, the rest being
masses of $d$ charged leptons and   $s$ Majorana neutrinos.
At $0<s\leq d$, there are also $s$ induced   Majorana masses,
$d-s$  neutrinos still remaining massless.\footnote{This reflects the
fact that in this case the rank of the neutrino mass matrix
is  $2s$.} At $s> d >0$, all the $d+s$ Majorana
neutrinos acquire masses. It is clear
that in contrast to the quark sector, the $CP$ violation
generally takes place  for more than  one singlet
neutrino at any $d>0$. The last  line in Table~1
illustrates  the extended $n$ family SM with one
right-handed neutrino per family.\footnote{Stress that in all the
tables the number of physical masses, chosen as independent ones, is
collectively that for both the charged
leptons and  neutrinos (Majorana or Dirac, depending on
context).}$^,$\footnote{Our
counting for the renormalizable $(d,s)_{\mbox{\scriptsize r}}$
extension, both at $s\leq d$ and $s>d$,  disagrees
with that in ref.~\cite{nm} (see remarks in section~4).}

In the  case $(d,0)_{\mbox{\scriptsize r}}$ one has $G=U(d)^2$, all
the neutrinos are massless  and the residual symmetry
increases up to $H=U(1)^d$ of the individual lepton numbers. Hence
the number of mixing angles, as well as that of physical phases, is
equal to zero.

\paragraph{(ii) Only  Dirac masses}
There is an important case of the SM renormalizable extension.  
Namely, the lepton number conservation  would forbid the Majorana
mass
terms, both the left- and right-handed. In the absence of these masses
the residual symmetry at $0<s\leq d$ would increase up to 
$H=U(1)$ of the total lepton number. In this case, designate it  
$(d,s)_{\mbox{\scriptsize D}}$, 
$2s$  degenerate in pairs Majorana neutrinos would
constitute $s$ massive Dirac ones, the rest being massless. Hence
there would be $s(2d-s-1)/2$  
mixing angles and $s(2d-s-1)/2-d+1$ phases~\cite{dir}. 
It follows, in particular, that at $s=d\equiv n$ for this reduced
type  of the $(n,n)_{\mbox{\scriptsize r}}$ extension one would get 
$2n$  masses, $n(n-1)/2$ mixing angles and   $(n-1)(n-2)/2$
phases in a complete analogy to the quark sector. 

The above results are not applicable at $s>d>0$. Here the number
of massive Dirac neutrinos saturates the
maximum allowed value $d$,   
the rest of $s-d$ Weyl neutrinos being massless. Hence the residual
symmetry would increase up to $H=U(s-d)\times U(1)$, so that the
number of  mixing angles would be $d(d-1)/2$ and the number of phases
$(d-1)(d-2)/2$. 
The results are summarized in Table~2 along with two particular
cases $(n,n-1)_{\mbox{\scriptsize D}}$ and $(n,n)_{\mbox{\scriptsize
D}}$ with equal numbers of mixing parameters. 
\begin{table}[htbp]
\paragraph{Table 2} Parameter counting for the  SM renormalizable
extension $(d,s)_{\mbox{\scriptsize D}}$  with only Dirac masses. 
The number of physical  masses is that of Dirac ones.

\vspace{1ex}
\begin{center}
\begin{tabular}{|c|c|c|}
\hline 
Couplings&Moduli&Phases\\
and symmetries&&\\
\hline
$Y^e$,  $Y^\nu $&$d^2+ds $&
$d^2+ds $\\ 
\hline
$G = U(d)^2\times U(s)$&$- d(d-1) - s(s-1)/2 $& 
$- d(d+1)-  s(s+1)/2 $\\
\hline
$H = U(1)$,\, $0<s\leq d $&$0$&$1$\\
\hline
$H =U(s-d)\times U(1)$ &$(s-d)(s-d-1)/2$&$(s-d)(s-d+1)/2+1$\\
$0<d< s$&&\\
\hline
${\cal M}_{ph}(d,s)_{\mbox{\scriptsize D}}$,\, $0<s\leq d $&
$s(2d-s-1)/2 +(d+s)$&$s(2d-s-1)/2-d+1$ \\
\hline
${\cal M}_{ph}(d,s)_{\mbox{\scriptsize D}}$,\, $0<d< s$&
$d(d-1)/2+2d$&$(d-1)(d-2)/2$\\
\hline
${\cal M}_{ph}(n,n-1)_{\mbox{\scriptsize D}}$\,, $n>1$&$
n(n-1)/2+(2n-1) $&$(n-1)(n-2)/2$\\
\hline
${\cal M}_{ph}(n,n)_{\mbox{\scriptsize D}}$\,, $n>0$&$
n(n-1)/2+2n $&$(n-1)(n-2)/2$\\
\hline
\end{tabular}
\end{center}
\end{table}

\section{Non-renormalizable extensions}
\paragraph{(i) Arbitrary case}
Let us now generalize the preceding
considerations to  the most exhaustive
Dirac-Majorana case with the left-handed  Majorana masses. 
The direct Majorana mass term for the
doublet neutrinos is excluded in the minimal SM by the symmetry
and
renormalizability requirements. But  in the extended
SM as a low energy effective theory, it could stem 
from the SM invariant operator  of the fifth dimension 
\begin{equation}\label{Eq:DeltaL}
- \Delta{\cal L}=\frac{1}{2\Lambda}\big(\phi^{C\dagger}
\tau_i\phi\big)
\big(\overline{l_R^{C}}h\,i\tau_2\tau_i l_L\big) +{\mbox
h.c.}\,,
\end{equation}
with $\tau_i$, $i=1,2,3$ being the Pauli matrices, $h$ being a
$d\times d$ symmetric constant matrix, $\Lambda\gg v$ being the
lepton number violating mass scale (supposedly of order of the
singlet Majorana masses) and $v$  being the Higgs vacuum
expectation value. The  above operator with the effective
isotriplet field
$\Delta_i=(1/\Lambda)\big(\phi^{C\dagger}\tau_i\phi\big)$ reflects
the oblique radiative corrections in the low
energy  Lagrangian produced by the physics beyond
the~SM.\footnote{Were the isotriplet  $\Delta_i$ be
considered as elementary 
in the framework of renormalizable extensions, it would change only
the  emerging Yukawa interactions
not affecting the mass and mixing matrices.} In the
unitary gauge, it yields the following mass and Yukawa  term 
\begin{equation}\label{eq:DeltaL'}
- \Delta{\cal L}= \frac{1}{2}\Big(
1+\frac{H}{v} \Big)^2\,
\overline{\nu_R^{C}}
\mu\, \nu_L +{\mbox h.c.} \,,
\end{equation}
with $\mu=hv^2/\Lambda $.

There is no nontrivial residual symmetry in this case  either,
$H=I$.
As for free parameters, phenomenological inclusion of such a mass
term  increases  the numbers of  moduli and phases by
$d(d+1)/2$ each. Of the extra moduli, $d$ ones are the Majorana
neutrino masses, the rest being  physical mixing angles. Hence, the
extension amounts to $d+s$ independent  neutrino  masses, $d(d
+2s-1)/2$ physical
mixing angles and the same number of phases~\cite{nm}.  Let us
designate this
general type of the SM extension as $(d,s)$, whether
$s\leq d$ or $s>d$. The parameter counting
for this non-renormalizable  extension of the SM is summarized in
Table~3. 

A  special case   without singlet neutrinos, i.e., $(d,0)$
extension, results in $d(d-1)/2$ mixing angles and the same number
of phases~\cite{n0}. Clearly, the $CP$
violation in the lepton sector becomes possible for two or more  
families without singlet neutrinos at all. On the other hand,  the
$(n,n)$ extension with $n$ complete families brings in $2n$
massive Majorana neutrinos with $n(3n-1)/2$ mixing angles and equal
number of phases~\cite{nn}. Hence, $CP$ violation might take
place here already for one complete family. 

\begin{table}[htbp]
\paragraph{Table 3} Parameter counting 
for the  SM non-renormalizable extension $(d,s)$
with  $d$ doublet  and $s$ singlet neutrinos. Symmetries are the same
as in Table~1.

\vspace{1ex}
\begin{center}
\begin{tabular}{|c|c|c|}
\hline 
Couplings&Moduli&Phases\\
\hline
$Y^e$,  $Y^\nu$, $M$   &$d^2+ds +s(s+1)/2$&
$d^2+ds +s(s+1)/2$\\ 
$\mu$&+$d(d+1)/2$&+$d(d+1)/2$\\
\hline
${\cal M}_{ph}(d,s)$ &$d(d +2s-1)/2 +(2d+s) $&$d(d
+2s-1)/2$\\
\hline
${\cal M}_{ph}(n,n)$&$n(3n-1)/2 +3n $&$n(3n-1)/2$\\
\hline
\end{tabular}
\end{center}
\end{table}

\paragraph{(ii) Only  Majorana masses}
Let us  consider a peculiar case of  the general extension above. 
In the absence of Yukawa couplings,  $Y^\nu\equiv 0$,
but at nonzero Majorana masses, both left- and right-handed,   the
residual symmetry  is still trivial ($H=I$) as in the
general case. But  now the doublet and singlet neutrino
sectors  completely disentangle from each other.  All the $d+s$
Majorana neutrinos  acquire masses, and we end  up with   $d(d-1)/2$
mixing angles and  the same number of phases for the doublet
neutrinos, without any mixing for the singlet
ones,  whether $s\leq d$ or $s>d$. Let us designate
this case  $(d,s)_{\mbox{\scriptsize M}}$. The results are
collected in Table~4.\footnote{Stress that the numbers of the physical
mixing angles and phases do not depend here  on $s$.  This is because
the right-handed neutrinos are sterile in the
case at hand,  and their mixing  matrix can  be chosen to be
unity in neglect of any other interactions.}
As for doublet neutrinos, this case formally corresponds to  
$(d,0)_{\mbox{\scriptsize M}}$  which, in turn, coincides with the
general one $(d,0)$. 
\begin{table}[htbp]
\vspace{-2ex}
\paragraph{Table 4} Parameter counting 
for the SM extension $(d,s)_{\mbox{\scriptsize M}}$  with
only Majorana  masses for neutrinos. Symmetries are the same as in
Table~1.
\vspace{1ex}
\begin{center}
\begin{tabular}{|c|c|c|}
\hline 
Couplings&Moduli&Phases\\
\hline
$Y^e$, $\mu$, $ M $&$d^2+d(d+1)/2 +s(s+1)/2 $&
$d^2+ d(d+1)/2 +s(s+1)/2$\\ 
\hline
${\cal M}_{ph}(d,s)_{\mbox{\scriptsize M}}$&
$d(d-1)/2 +(2d+s)$&$d(d-1)/2$ \\
\hline
\end{tabular}
\end{center}
\end{table} 

\section{Remarks}

We would like to clarify some discrepancies for the renormalizable
$(d,s)_{\mbox{\scriptsize r}}$ extension 
between our counting in the weak basis 
and the one in the mass basis~\cite{nm}. In the mass basis,
an explicit new feature of the $(d,s)_{\mbox{\scriptsize r}}$
extension, compared to $(d,s)$ one, is the appearance of
the additional symmetry $U(d-s)$, $d\geq s$, due to  $d-s$  neutrinos
being massless. As a result, it is stated in the cited paper
that the mixing
matrix for the $(d,s)_{\mbox{\scriptsize r}}$  extension could be
obtained from  the corresponding general
matrix just by deleting in the latter  $(d-s)^2$ spurious parameters
corresponding to $U(d-s)$. We would like to remark that this procedure
though being applicable does not fix the number of independent
parameters and
generally overestimates the number of the actual ones.

To illustrate, note that it would follow from
the prescription~\cite{nm}, e.g., that at $d=s\equiv n$ both  
$(n,n)_{\mbox{\scriptsize r}}$ and $(n,n)$ extensions  would have
the same numbers of mixing angles, as well as  phases,
$n(3n-1)/2$, in addition to $3n$ masses. On the other hand,  an
arbitrary square complex matrix $Y$ can be uniquely
written as a unitary matrix times a positive-definite Hermitian
one, and  a complex symmetric matrix $M$ can be uniquely
decomposed in terms of a unitary matrix $V$ and a positive-definite
diagonal one,  $M=V^T M_{diag} V$. This means that with account
for the global symmetry $G$ we could start in the
$(n,n)_{\mbox{\scriptsize r}}$ extension
by choosing from the very beginning 
the Yukawa matrices $Y^e$ and $Y^\nu$ as positive-definite Hermitian
matrices and $M$ as a positive definite diagonal one. As we have
exhausted thus all the symmetry $G$ and there is no nontrivial
residual subgroup $H$, this set of parameters is the independent
physical one. It contains $n(n+1)+n$ 
moduli and $n(n-1)$ phases.
This completely agrees with Table~1 and 
is  clearly less compared to ref.~\cite{nm}.

We trace the origin of the  discrepancy between the countings to
the constraint $\mu=0$ in eq.~(\ref{eq:DeltaL'}). In passing from 
$(d,s)$  extension to the $(d,s)_{\mbox{\scriptsize r}}$ one, it
restricts $d(d+1)/2$  phases and  the same number of physical moduli.
In this, $d$ of the  conditions on moduli serve to determine the
induced Majorana masses.
Altogether, this leaves $s$ independent Majorana masses, $sd$  mixing
angles and
$d(s-1)$ phases. At $0<s\leq d$, there are $s$ induced nonzero masses,
$d-s$ neutrinos remaining massless with necessity. 
As a consequence of the inborn masslessness for $d-s$ neutrinos, the
stated constraint supersedes here those gained from the $U(d-s)$
symmetry. 
E.g., according to prescription~\cite{nm} the extension 
$(d,1)_{\mbox{\scriptsize r}}$ formally corresponds
to $2d-1$ mixing angles and $d$ phases, but explicit construction
shows that  there are actually just
$d$  mixing angles, all of them being independent, and no
phases at all. Especially clear the above
constraint works at $s> d$ when there appear no massless
neutrinos and there is nothing to delete by the related
transformations. Nevertheless, 
the counting of  parameters at $s>d$ for the
$(d,s)_{\mbox{\scriptsize r}}$
extension  proves to be not the same as for the $(d,s)$ one.

\section{Conclusion}

The parameter counting in the weak basis is complementary to that in
the mass basis. It allows one to gain clear insight into the
independent physical parameter content of the SM extensions, 
both renormalizable and non-renormalizable, 
with arbitrary numbers of the singlet and  left-handed doublet
neutrinos. 

The author is grateful to V.V.\ Kabachenko and L.B.\ Okun for very
useful discussions.


\begin{thebibliography}{**}

\bibitem{koba}
N.\ Cabibbo, Phys.\ Rev.\ Lett.\ {\bf 10} (1963) 531;
M.\ Kobayashi and T.\ Mas\-ka\-wa, Prog.\ Theor.\ Phys.\ {\bf 49}
(1972) 652; 
L.\ Maiani,  Phys.\ Lett.\ {\bf 62B} (1976) 183. 

\bibitem{ponte}
B.\ Pontecorvo, ZhETF {\bf 34} (1958) 247 [Sov.\ Phys.\ JETP {\bf 7}
(1958) 172]; {\it ibid.} {\bf 53} (1967) 1717 [Sov.\ Phys.\ JETP {\bf
26} (1968) 984]; Z.~Maki, M.\ Nakagawa and S.\ Sakata, Prog.\ Theor.\
Phys.\ {\bf 28} (1962) 870; 
V.\ Gribov and B.\ Pontecorvo, Phys.\ Lett.\ {\bf 28B} (1969) 493;
S.M.~Bilen\-ky and B.\ Pontecorvo, Phys.\ Rep.\ {\bf 41C} (1978) 225.

\bibitem{now}
S.M.\ Bilenky {\it et al.}, Summary of the NOW'98 Phenomenology
Working Group, hep-ph/9906251.

\bibitem{n0}
J.~Schechter and J.W.F.\ Valle, Phys.\ Rev.\ {\bf D23} (1981) 1666;
M.~Doi, T.~Kotani, H.\ Nishiura, K. Okuda and E.\
Takasugi, Phys.\ Lett.\ {\bf 102B} (1981) 323; 
J.~Bernabeu and P.~Pascual, Nucl.\ Phys.\ {\bf B228} (1983) 21;
P.J.\  O'Donnell and U.\ Sarkar, Phys.\ Rev.\ {\bf D52} (1995) 1720,
hep-ph/9305338.

\bibitem{nn}
I.Yu.\ Kobzarev, B.V.\ Martemyanov, L.B.\ Okun and M.G.\ Schepkin,
Yad.\ Fiz.\ {\bf 32} (1980) 1590;
S.M.\ Bilenky, J.\ Hosek and S.V.~Petcov, Phys.\ Lett.\ {\bf 94B}
(1980) 495; V.~Barger, P.~Langacker, J.P.~Leveille and S.~Pakvasa,
Phys.\ Rev.\ Lett.\ {\bf 45} (1980) 692. 
 
\bibitem{nm}
J.\ Schechter and J.W.F.\ Valle, Phys.\ Rev.\ {\bf D22} (1980)
2227.

\bibitem{dir}
J.\ Donoghue, Phys.\ Rev.\ {\bf D18} (1978) 1632;
J.\ Schechter and J.W.F.~Val\-le, {\it ibid.} {\bf D21} (1980) 309.

\bibitem{santa}
A.\ Santamaria, Phys.\ Lett.\ {\bf 305B} (1993) 90, hep-ph/9302301.

\bibitem{4f}
Yu.F.\ Pirogov and O.V.\ Zenin, Eur.\ Phys.\ J.\ {\bf C10} (1999)
629, hep-ph/9808396.

\end{thebibliography}
\end{document}